\documentclass[reprint,prd,amsmath,amssymb,aps,lengthcheck,floatfix,superscriptaddress]{revtex4-1}
\usepackage{balance}
\usepackage{graphicx}
\usepackage{dcolumn}
\usepackage{bm}
\usepackage{hyperref}
\usepackage[utf8]{inputenc}
\usepackage{amsmath}
\usepackage{multirow}
\usepackage{soul}
\usepackage{float}
\usepackage{graphicx}
\usepackage{times}
\usepackage[normalem]{ulem}
\usepackage{color}
\usepackage{cellspace}
\usepackage{lipsum}

\newcommand{\be}{\begin{equation}}
\newcommand{\ee}{\end{equation}}
\newcommand{\bea}{\begin{eqnarray}}
\newcommand{\eea}{\end{eqnarray}}

\newcommand{\comment}[1]{}
\renewcommand\sout{\bgroup \color{red} \ULdepth=-.5ex \ULset}
\usepackage{xcolor}

\def\simge{\mathrel{\rlap{\raise 0.511ex
     \hbox{$>$}}{\lower 0.511ex \hbox{$\sim$}}}}
\def\simle{\mathrel{\rlap{\raise 0.511ex
      \hbox{$<$}}{\lower 0.511ex \hbox{$\sim$}}}}

\usepackage{orcidlink}

\begin{document}

\title{Astrophysics and Nuclear Physics Informed Interactions in Dense Matter: Inclusion of PSR J0437-4715}

\author{Tuhin Malik \orcidlink{0000-0003-2633-5821}}
\affiliation{CFisUC, Department of Physics, University of Coimbra, P-3004 - 516 Coimbra, Portugal}

\author{Veronica Dexheimer \orcidlink{0000-0001-5578-2626}}
\affiliation{Department of Physics, Kent State University, Kent, OH 44243, USA}

\author{Constan\c ca Provid\^encia \orcidlink{0000-0001-6464-8023}}
\affiliation{CFisUC, Department of Physics, University of Coimbra, P-3004 - 516 Coimbra, Portugal}

\date{\today}
\begin{abstract}
We investigate how vector-isoscalar and {vector}-isovector interactions can be determined within the density regime of neutron stars (NSs), while fulfilling nuclear and astrophysics constrains. We make use of the Chiral Mean Field (CMF) model, a SU(3) nonlinear realization of the sigma model within the mean-field approximation, for the first time within a Bayesian analysis framework. We show that neutron-matter $\chi$EFT constraints at low density are only satisfied if the vector-isovector mixed interaction term is included, e.g., a $\omega^2\rho^2$ term. We also show the behavior of the model with respect to the conformal limit. We demonstrate that the CMF model is able to predict a value for the parameter $d_c$ related to the trace anomaly and its derivative takes values below 0.2 above four times saturation density within a hadronic {version of the} model that does not include { hyperons or} a phase transition to deconfined matter. We compare these effects with results from other (non-chiral) Relativistic Mean Field models to assess how different approaches to incorporating the same physical constraints affect predictions of NS properties and dense matter equations of state. We also include data from the gravitation wave event GW230529 detected  by the LIGO-Virgo-Kagra collaboration and  the most recent radius measurement of PSR J0437-4715 from the NASA NICER mission. Our analysis reveals that this new NICER measurement leads to an average reduction of approximately $\sim 0.1$ km radius in the posterior of the NS mass-radius relationship.
\end{abstract}

\maketitle

\textbf{\textit{Introduction-}}
 Neutron stars (NSs), which are remnants of supernova explosions, serve as celestial markers that shed light on the boundaries of physical principles \cite{Oertel:2016bki}. These objects, known for their extremely high density, $\sim$ 4-5 times nuclear saturation density {$\rho_0$}, provide a natural setting for investigating the fundamental characteristics of matter in extreme environments \cite{book.Glendenning1996,Burgio:2021vgk}. The equation of state (EoS) of NS matter, crucial for unraveling the mysteries surrounding these cosmic entities, explains the relationships among their physical attributes such as density, pressure, and temperature. Theoretical modeling of EoS plays a vital role in understanding the inner composition and behavior of NSs \cite{book.Haensel2007,Tan:2020ics,Tan:2021nat, Marczenko:2022cwq, Altiparmak:2022bke}.

We examine the EoS of NS matter using the Chiral Mean Field (CMF) model {within} the framework of the SU(3) nonlinear sigma model, within the mean-field approximation. This approach enables a detailed examination of vector and isovector interactions, which are crucial factors in describing the repulsion component of the strong force and consequently shaping the EoS of NSs. The model incorporates meson exchange interactions to describe hadronic systems, with a keen focus on maintaining chiral invariance \cite{Papazoglou:1998vr,Bonanno:2008tt}, an important characteristic of Quantum chromodynamics (QCD). This ensures that particle masses, emerging from interactions within the medium, approach zero under extreme conditions, aligning with the properties of high-density and/or high-temperature environments. {In this model we can easily explore various self-interactions among vector-isoscalar  and -isovector mesons, while adhering to chiral invariance principles. These have already shown to be extremely important on predicted microscopic and macroscopic observables \cite{Steiner:2012rk,Dutra:2014qga,Providencia:2012rx,Kubis:1997ew,Liu:2001iz,Gomes:2014aka,Menezes:2004vr}}. Through these considerations, the updated model aims to align closely with the latest empirical data and theoretical insights into nuclear physics, thereby enhancing its predictive power and relevance to current astrophysical and nuclear physics research.

In this work, we utilize for the first time advanced parameter estimation techniques using Bayesian inference to investigate the CMF model in the setting of high-density environments commonly found in NSs. Bayesian inference, a widely recognized statistical technique for improving parameter estimates based on prior information, offers a statistical approach to understanding the EoS of NS matter \cite{Imam:2021dbe,Malik:2022ilb,Coughlin:2019kqf,Wesolowski:2015fqa,Furnstahl:2015rha,Ashton2019,Landry:2020vaw,Huang2023:inpress,Char:2023fue,Imam:2024gfh}. By employing this approach, our goal is to methodically investigate the impacts of vector and isovector interactions, ultimately advancing our understanding of the underlying physics that governs the interior of  NSs.

A key focus in this letter is to highlight the importance of vector self-interaction and mixed terms, specifically the $\omega^4$ and $\omega^2 \rho^2$ terms, and {the significant impact of the last one}  on matching pure neutron-matter requirements at low densities. These terms are essential components of the model and affect the density variation of the symmetry energy and the speed of sound. {   It is the density dependence of the symmetry energy that determines the onset of fast cooling processes, such as the nucleonic direct Urca processes, and also influences the radius {  and tidal deformability} of low to medium mass NSs \cite{Cavagnoli:2011ft,Dexheimer:2018dhb}.} 
Our results highlight the essential nature of including these terms correctly to align with the observed behavior of NSs. Here, we focus solely on nucleonic degrees of freedom and investigate how chiral symmetry is restored within astrophysics and nuclear physics informed interactions in dense matter. 
{ {While there is a possibility that hyperons, spin 3/2 baryons, and deconfined quarks appear in {dense} matter in the regime relevant for astrophysics, there is also a chance that they appear at larger densities in the zero-temperature limit. Ab-initio calculations have shown that when including three-body forces in their calculations, hyperons might not be present in NSs or appear in very small amounts \cite{Gerstung:2020ktv,Tong:2024egi} and the quarkyonic picture predicts that quark deconfinement takes place at much larger densities than chiral symmetry restoration at low temperatures (see Fig.~1 of Ref.~\cite{McLerran:2007qj}).}}
Our goal is to perform an inference analysis using a chiral effective field theory ($\chi$EFT) \cite{Huth:2021bsp} informed prior, and apply constraints derived from perturbative QCD (pQCD) \cite{Kurkela:2014vha}, and astrophysical observations including {  the latest} NASA's Neutron star Interior Composition Explorer (NICER) \cite{Riley:2019yda,Riley:2021pdl,Miller:2019cac,Miller:2021qha,Choudhury:2024xbk} and LIGO-VIRGO Collaboration (LVC) data \cite{LIGOScientific:2018cki}, see also \cite{Huth:2021bsp,Tsang2024}.

\textbf{\textit{Microscopic EoS-}} For this work we use the nucleonic version of the model (including only protons and neutrons)~\cite{Dexheimer:2008ax} and relax all the vector self-interaction and mixed-interaction parameters, which are now freely varied, {together with the couplings $g_\omega$ and $g_\rho$ of the $\omega$ and $\rho$ mesons to the nucleon,} {and determined within an inference calculation}. The scalar interactions remain fitted to reproduce nucleon masses, as well as meson masses and decay constants (see Ref.~\cite{Kumar:2024owe} for details). {  To describe NS matter, we also include electrons in $\beta$-equilibrium and fulfilling charge neutrality.}
The CMF Lagrangian density can be written as:
$
\mathcal{L} = \mathcal{L}_{\rm{Kin}}+\mathcal{L}_{\rm{Int}}+\mathcal{L}_{\rm{Self}}+\mathcal{L}_{\rm{SB}}\,,
$
with terms for the kinetic energy, interactions between nucleons and vector and scalar mesons, self interactions of scalar and vector mesons, and explicit chiral symmetry breaking. As there is no strangeness included in the system, we do not include mesons with hidden strangeness. This leaves the vector-isoscalar $\omega$ and vector-isovector $\rho$, the scalar-isoscalar $\sigma$ and scalar-isovector $\delta$ to mediate the strong force. The isovector mesons affect isospin-asymmetric matter and, thus, are important to describe NSs.

The effective masses for the nucleons $i$ are generated by the scalar mesons,
$
M_{i}^*=g_{i\sigma}\sigma+g_{i\delta}\tau_3\delta+\Delta m_i
$,
with a small explicit mass term $\Delta m_i$, whose possible values we explore.
The main different chiral invariant possibilities for the vector self-interaction terms are {  (see Ref.~\cite{Gasiorowicz:1969kn} for more details on chiral invariant terms)}:
$ 
2{\rm{Tr}}(V^4),\,
\bigl(({\rm{Tr}}(V^2)\bigl)^2,\,
\bigl(({\rm{Tr}}(V)\bigl)^4/4, \label{vector}
$ 
where $V$ is the vector-meson multiplet matrix, which reduces to a diagonal matrix in the mean-field approximation:
$
V = {\rm{diag}} \bigl((\omega + \rho)/\sqrt{2}, (\omega - \rho)/\sqrt{2}, 0 \bigl) \,.
$
{All other chiral invariants can be derived from these (in the absence of strangeness). In the presence of strangeness there are two more chiral invariant couplings, but they were never studied in detailed because they did not seem to produce physical results. In the absence of strangeness, these 2 additional couplings can be written using the couplings we discuss.}
The different self-interaction terms of the vector mesons shown above correspond, respectively, to the coupling schemes
C1:
$\mathcal{L}^{\rm{Self}}_{\rm{vec}} = g_{4,1}(\omega^4 + 6\omega^2\rho^2 + \rho^4)$,
C3:
$\mathcal{L}^{\rm{Self}}_{\rm{vec}} = g_{4,3}(\omega^4 + 2\omega^2\rho^2 + \rho^4)$, and 
C4:
$\mathcal{L}^{\rm{Self}}_{\rm{vec}} = g_{4,4}(\omega^4) $. 
{The additional coupling scheme C2:
$\mathcal{L}^{\rm{Self}}_{\rm{vec}} = g_{4,2}(\omega^4 + \rho^4)$} is a linear combination of the others.
Note that in the past we used either only one of the {coupling schemes} {C1-C4} \cite{Dexheimer:2015qha}, or we used C4 and added non-chiral invariant terms  \cite{Dexheimer:2018dhb,Dexheimer:2020rlp, Kumar:2024owe}, always including the meson with hidden strangeness $\phi$. 

{Here we preserve chiral invariance and, instead of adding terms, we study particular combinations of the {coupling schemes} above that allow us to 1) isolate each one of the three independent terms:
\begin{itemize}
\item {\bf{x}}: $\mathcal{L}^{\rm{Self}}_{\rm{vec}} =$ x$\rho^2\omega^2$\ ;
\item {\bf{y}}: $\mathcal{L}^{\rm{Self}}_{\rm{vec}} =$ y$\rho^4$\ ;
\item {\bf{z}}: $\mathcal{L}^{\rm{Self}}_{\rm{vec}} =$ z$\omega^4$\ ;
\end{itemize}
2) consider the combination of two terms:  
\begin{itemize}
\item 
{\bf{xz}}: $\mathcal{L}^{\rm{Self}}_{\rm{vec}} =$ x$\rho^2\omega^2$ + z$\omega^4$\ ;
\end{itemize}
and 3) consider a combination of the three terms:
\begin{itemize}
\item 
{\bf{xyz}}: $\mathcal{L}^{\rm{Self}}_{\rm{vec}} =$ x$\rho^2\omega^2$ + y$\rho^4$ + z$\omega^4$\ .
\end{itemize}}
{The correspondence between x, y, and z and the $g_4$ couplings is trivial. E.g., for case 3), one could write that x=$2g_{4.3}$, y=$g_{4.2}+g_{4.3}$, and z=$g_{4.2}+g_{4.3}+g_{4.4}$.}

\textbf{\textit{Bayesian Likelihood-}}
The parameters of the CMF model are determined {here} through the Bayesian Inference framework. We enforce minimal nuclear saturation properties (NMP): (i) the saturation density $\rho_0 = 0.16 \pm 0.005$ fm$^{-3}$, binding energy per nucleon $\epsilon_0 = -16.1 \pm 0.2$~MeV, and symmetry energy $J_0 = 30 \pm 2$~MeV at saturation \cite{MUSES:2023hyz}, (ii) constraints on low-density neutron matter from various $\chi$EFT calculations regarding the energy per particle at densities $0.05$, $0.1$, $0.15$, and $0.20$ fm$^{-3}$ \cite{Huth:2021bsp}, (iii) constraints derived from pQCD at seven times {$\rho_0$} for the highest renormalizable scale X=4 \cite{Komoltsev:2021jzg}, (iv) astrophysical constraints such as mass-radius measurements from PSR J0030+0451 \cite{Vinciguerra:2023qxq,Riley:2019yda,Miller:2019cac} and PSR J0740+6620 \cite{Salmi:2022cgy,Riley:2021pdl,Miller:2021qha}, and tidal deformability from GW170817 \cite{LIGOScientific:2018cki}. We also discuss recent mass-radius NICER results for PSR J0437-4715 \cite{Choudhury:2024xbk}.

Bayesian probability functions are defined as the probability of observation data specified in the statistical model. {The posterior distributions of the model parameters $\theta$ in Bayes’ theorem are expressed as
$
P(\bm{\theta} |D ) ={{\mathcal L } (D|\bm{\theta}) P(\bm {\theta })}/{\mathcal Z},\label{eq:bt}
$
where the model parameters $\bm{\theta}$ are  the vector interaction couplings,  $D$ denotes  the fit data which are defined below,  $P(\bm {\theta })$ is the prior for the  model parameters and $\mathcal Z$ is the evidence.}
 The probability of a set of constraints and the posterior distribution of astrophysical observations can be calculated as follows:

(i) {nuclear matter properties (NMP): with \(D_{\text{NMP}} \pm \sigma\) representing the desired value or the value to be fitted (where the data follows a symmetric Gaussian distribution) and \(D(\theta)\) representing the predicted value for that quantity for a given parameter set, the likelihood is given by:
\begin{equation}
{\mathcal L}({\mathcal D_{\rm NMP} | }\theta) = \frac{1}{\sqrt{2\pi\sigma^2}} \exp(\frac{-(D(\theta)-D_{\rm NMP})^2}{2\sigma^2})= \mathcal{L}^{\rm NMP}
\end{equation}}
{(ii) The Low density PNM constraints for $\chi$EFT: a super-Gaussian type likelihood as 
\begin{equation}
\mathcal{L}^{\rm PNM} (\epsilon_{\chi\text{EFT}, i} | \theta ) = \frac{1}{2 \sigma_i} \cdot \frac{1}{\exp\left(\frac{\left| \epsilon_{\chi\text{EFT}, i} - \epsilon_{{\rm PNM}, i}(\theta) \right| - \sigma_i}{p}\right) + 1}
\end{equation}
Here, $\sigma_i$ denotes the uncertainty in the PNM data at the $i$-th point, with $p$ set to 0.015.} 

(iii)  perturbative Quantum Chromodynamics (pQCD): with $d_{\rm pQCD}$ representing a constant probability distribution throughout the area enclosed in the energy density and pressure plane at 7 times {$\rho_0$} calculated for the renormalizable scale X=4, the likelihood is given by:
\bea
{\mathcal L}({d_{\rm pQCD} | }\theta) &=P(d_{\rm pQCD}|\theta)=& \mathcal{L}^{\rm pQCD}\ ,
\eea
where $P(d_{\rm pQCD}|\theta) = 1$ if it is within $d_{\rm pQCD}$; otherwise zero;

(iv) gravitational-wave (GW) observation: information about EoS parameters comes from the masses $m_1, m_2$ of the two binary components and the corresponding tidal deformabilities, $\Lambda_1$ and $\Lambda_2$. In this case, the likelihood is \cite{LVKcollaboration}:
\begin{align}
    P(d_{\mathrm{GW}}|\mathrm{EoS}) = \int^{M_{\mathrm{max} }}_{M_\mathrm{min}}dm_1 \int^{m_1}_{M_\mathrm{min}} dm_2 P(m_1,m_2|\mathrm{EoS})   \nonumber \\
    \times P(d_{\mathrm{GW}} | m_1, m_2, \Lambda_1 (m_1,\mathrm{EoS}), \Lambda_2 (m_2,\mathrm{EoS})) 
    =\mathcal{L}^{\rm GW} \ , 
    \label{eq:GW-evidence}
\end{align}
where P(m$|$EoS) ~\cite{Agathos_2015,Wysocki-2020,Landry:2020vaw,Biswas:2020puz} can be written as:
\begin{equation}
    P(m|\rm{EoS}) = \left\{ \begin{matrix} \frac{1}{M_\mathrm{max} - M_\mathrm{min}} & \text{ if  } M_\mathrm{min} \leq m \leq M_\mathrm{max}\ , \\ 0 & \text{otherwise.} & \end{matrix} \right .
\end{equation}
{  Here, $M_{\rm min}$ is 1 $M_\odot$, and $M_{\rm max}$ represents the maximum mass of a NS for the given equation of state (EOS) \cite{Landry:2020vaw}.}

(v) X-ray observation (NICER) : these give simultaneous mass and radius measurements of NSs. Therefore, the corresponding evidence takes the following form,
\begin{align}
    P(d_{\rm X-ray}|\mathrm{EoS}) = \int^{M_{\mathrm{max} }}_{M_{\mathrm{min} }} dm P(m|\mathrm{EoS}) \nonumber \\ \times
    P(d_{\rm X-ray} | m, R (m, \mathrm{EoS})) 
    = \mathcal{L}^{\rm NICER}\ .
\end{align}
The final likelihood for the calculation is then given by
\begin{equation}
    \mathcal{L} = 
    \mathcal{L}^{\rm NMP}
    \mathcal{L}^{\rm PNM}
    \mathcal{L}^{\rm pQCD}
    \mathcal{L}^{\rm GW}
    \mathcal{L}^{\rm NICER I}
    \mathcal{L}^{\rm NICER II}\
    \mathcal{L}^{\rm NICER III}\,.
    \label{eq:finllhd}
\end{equation}
NICER I and II  correspond to the mass-radius measurements of PSR J0030+0451 \footnote{~\href{https://zenodo.org/records/4697625}{https://zenodo.org/records/4697625}} and PSR J0740+6620 \footnote{~\href{https://zenodo.org/records/4670689}{https://zenodo.org/records/4670689}}, respectively.
The NICER III dataset consists of the recent mass-radius measurements for PSR J0437-4715 \footnote{~\href{https://doi.org/10.5281/zenodo.10886504}{https://doi.org/10.5281/zenodo.10886504}}.

\textbf{\textit{Results-}}
Within the CMF model, we explore various chiral invariant scenarios for the vector non-linear terms within the Bayesian Inference framework, supported by the selection of nuclear and astrophysical constraints already discussed. We perform {five} different identical inference analyzes using different linear combinations of the C1-C4 terms {but varying directly the x, y, and z coupling constants. First, we include only one term and vary its respective coupling. Then we include combinations of two terms (varying their respective couplings xy) and finally of three terms (varying their respective couplings x, y, and z)}.
{Note that the terms x, and z  have a direct effect on, respectively, the symmetry energy and on the softness of the EoS at large densities \cite{Mueller:1996pm}. Their effect on the NS mass radius properties is to decrease the radius of low-mass stars ({with a positive x}) and to decrease the maximum mass and the radius of high mass stars (z with a positive coupling) \cite{Carriere:2002bx,Fattoyev:2010rx,Cavagnoli:2011ft}. We have also considered the {term} combination xy {(x$\omega^2\rho^2+$y$\rho^4$)}, however the results were almost coincident with the ones obtained with x, and, therefore, in the following this combination is not considered. }

\begin{figure}
    \centering
    \includegraphics[width=0.8\linewidth]{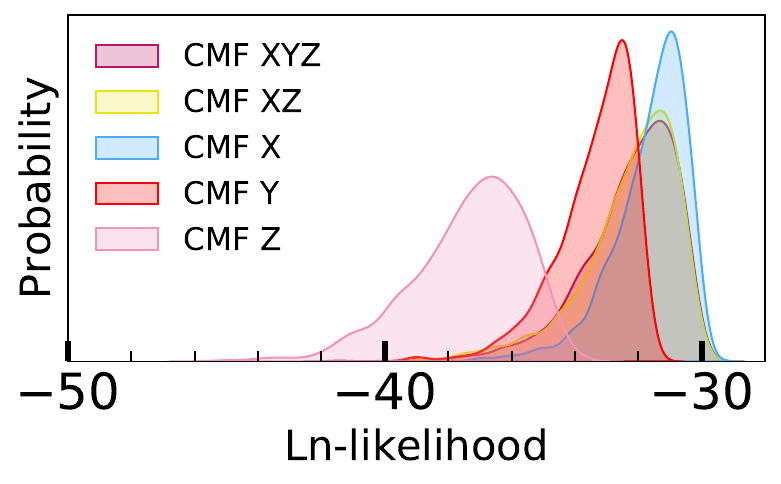}
    \caption{The probability distributions of the value of the log likelihood for the three sets analyzed including in each one all linear combinations of the different chiral invariant terms.}
    \label{fig:lnlike}
\end{figure}

{In Fig. \ref{fig:lnlike}, the probability distributions of the value of the log likelihood have been plotted for all sets.
The model with the highest likelihood is considered the best fit to the data.   Clearly, set z is the one that reproduces the data the worst, followed by set y. Set x has the greatest likelihood given the constraints imposed. The other two sets, xz and xyz, give similar results, indicating that the role of y is negligible. To determine the best supported model, we have calculated the Bayesian evidence for all sets and found the natural logarithm of Bayes factor $\ln K_{\rm{xyz},i}=\ln ({\cal Z}_{\rm{xyz}})-\ln({\cal Z}_{i})$ of model xyz with respect to model $i$,  where ${\cal Z}_i$ is the evidence of model $i$. We have obtained $\ln K_{\rm{xyz,xz}}=0.05$, $\ln K_{\rm{xyz,x}}=-0.73$, $\ln K_{\rm{xyz,y}}=3.4$, $\ln K_{\rm{xyz,z}}=6.09$, showing that there is a strong evidence of model xyz with respect to models y and z, but no large difference with respect to models x and xz.
}

\begin{figure*}
    \centering
    \includegraphics[width=0.33\linewidth]{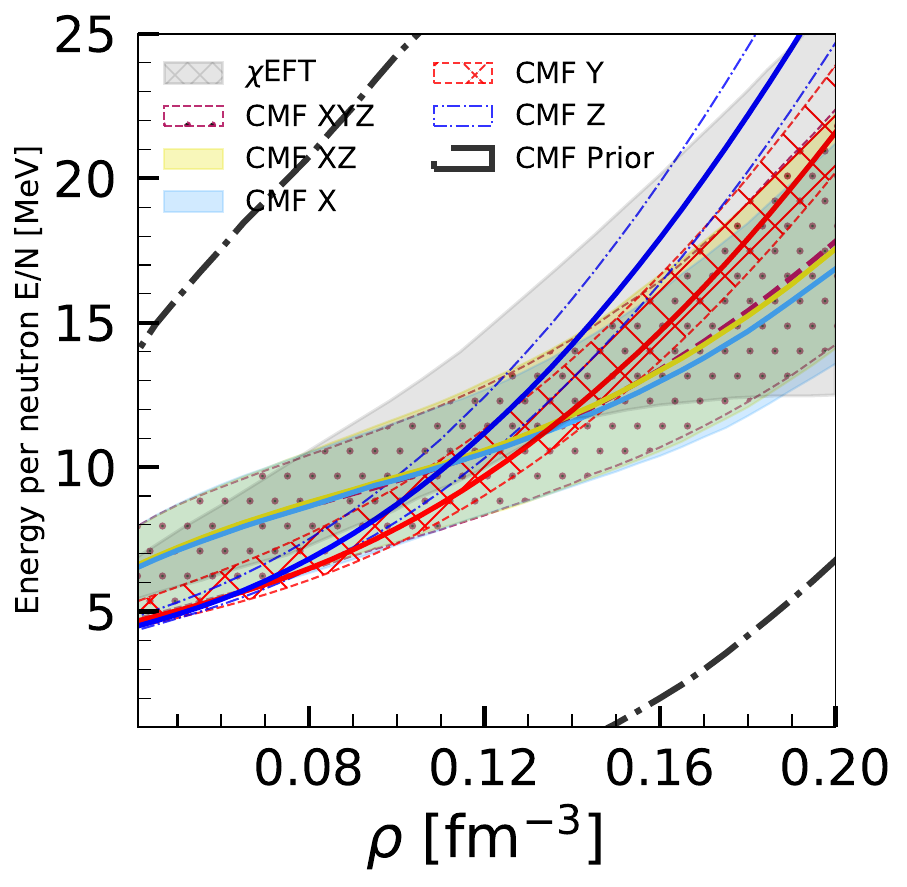}
    \includegraphics[width=0.33\linewidth]{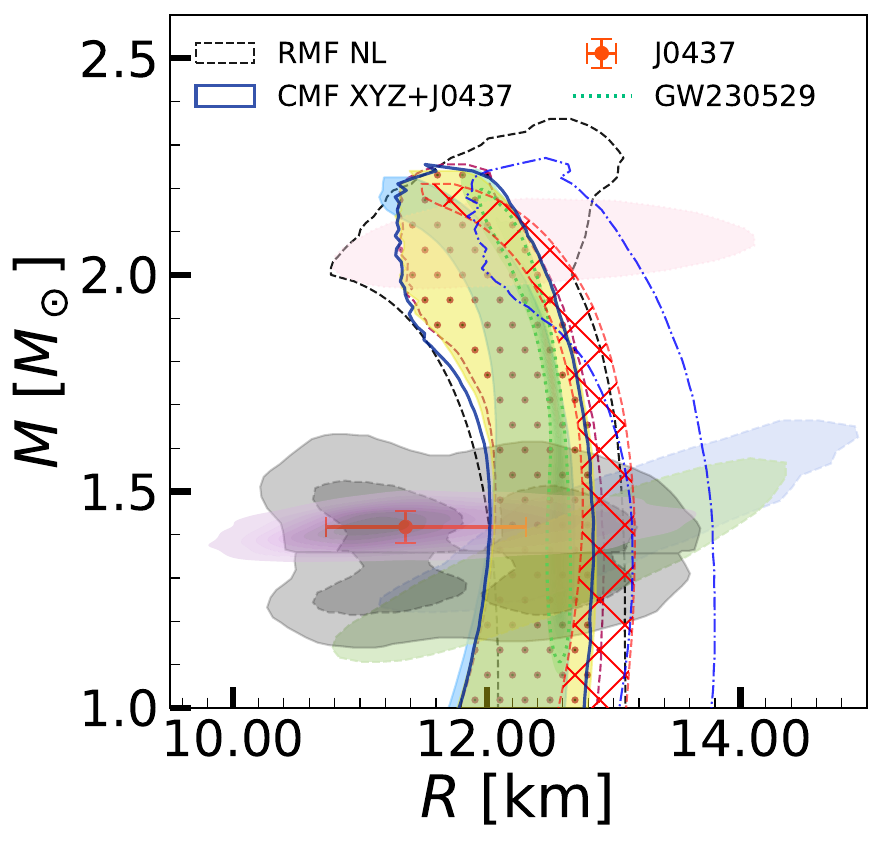}
    \includegraphics[width=0.33\linewidth]{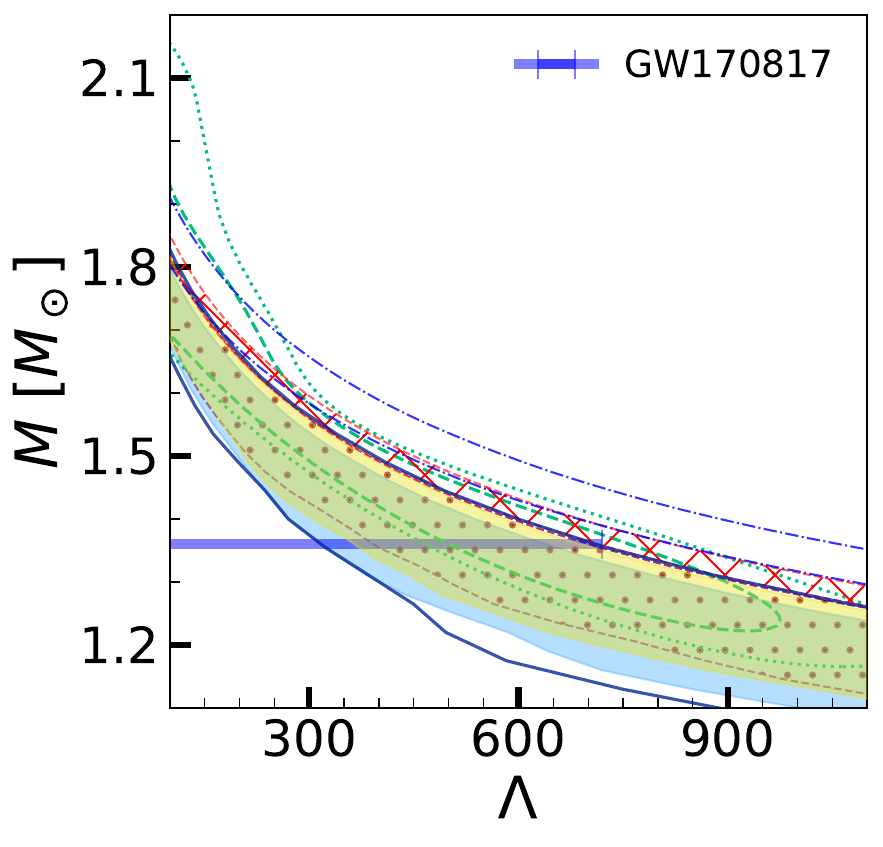}
    \caption{Posteriors obtained within the 90\% CI for all sets of CMF interactions. Left: Neutron-matter energy per particle as a function of density. The gray band includes constraints from various many-body calculations using $\chi$EFT interactions \cite{Huth:2021bsp}. The black dash-dotted line represents the xyz prior. The x, xz, and xyz bands are nearly coincident at low densities and differ only at higher densities, as seen by the medians.  Center: mass radius region. Gray lines represent {boundaries for} the binary components of GW170817, including their 90\% and 50\% CI. The $1 \sigma$ (68\%) CI for the 2D posterior mass-radius distribution for the millisecond pulsars PSR J0030+0451 (in pastel blue and soft green) \cite{Riley:2019yda, Miller:2019cac} and PSR J0740+6620 (in blush pink) \cite{Riley:2021pdl, Miller:2021qha} from the NICER X-ray data are shown. The black thin dashed region represents the 90\% CI obtained in a previous study using RMF and nonlinear mesonic interaction \cite{Malik:2023mnx}. The green dotted region indicates the inferred 90\% CI of GW230529 with BSK24 \cite{LIGO-P2300352-v9}. The recent NASA NICER data for PSR J0437-4715 are also shown \cite{Choudhury:2024xbk} {(orange dot with error bars and lilas region)}. The black thick solid line delineates the 90\% CI when data from PSR J0437-4715 are included in the inference using the xyz prior.  Right: Dimensionless tidal deformability for NS masses. The bar represents tidal deformability constraints at 1.36 \(M_\odot\) \cite{LIGOScientific:2018cki}. The green dashed (dotted) regions are the posterior $1\sigma$ ($2\sigma$) obtained for the secondary component in GW230529 using BSK24 \cite{LIGO-P2300352-v9}.}
    \label{fig:pnm}
\end{figure*}

In Fig. \ref{fig:pnm}, we plot the results of our Bayesian inference calculation, in particular, the posteriors for the neutron matter energy per particle as a function of the baryonic density (left), for the mass-radius distributions (middle) and for the mass tidal deformability distributions (right) acquired within a 90\% confidence interval (CI) for all different sets. 
In the left panel  the prediction from several $\chi$EFT  calculations as given in \cite{Huth:2021bsp} is also shown in hatched gray. 
{The sets x, xz and xyz give similar results and reproduce the low density results of $\chi$EFT. It is the presence of the term x that is responsible for the good agreement. The terms y and z alone do not reproduce the low density constraints.}
Note, however, that there is no credible interval associated with the $\chi$EFT data. {In this figure, the black dot-dashed line represents the prior of the xyz model. These terms have little effect on the properties of symmetric nuclear matter at saturation.}

\begin{figure}
    \centering
    \includegraphics[width=1.\linewidth]{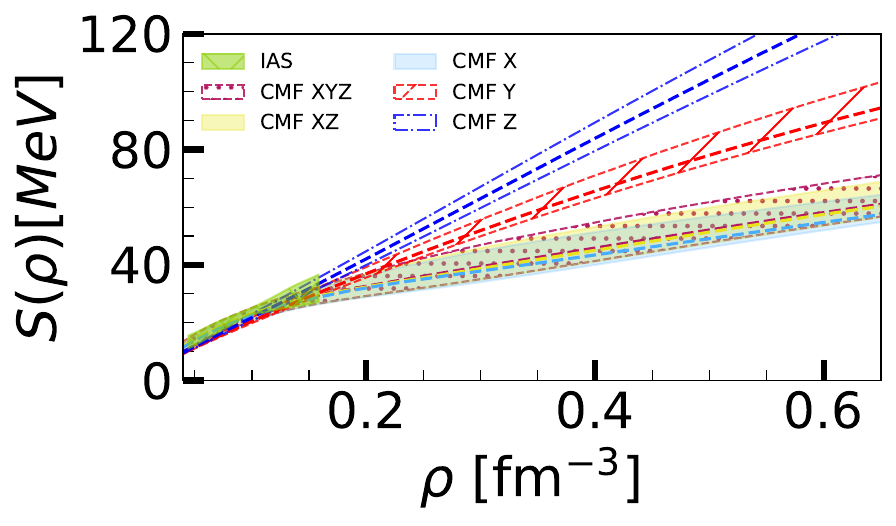}
    \caption{Symmetry energy posterior with respect to baryon density obtained within the 90\% CI for the five distinct groups of CMF instances under study.
    We also include the constraints from IAS \cite{Danielewicz:2013upa}.}
    \label{fig:symeng}
\end{figure}

Solving the Tolmann-Oppenheimer-Volkoff equations \cite{TOV1,Oppenheimer:1939ne} to determine the mass and radius of static spherical NSs, we have obtained the mass-radius distributions shown in the middle panel of Fig. \ref{fig:pnm} for the {five} sets. {The crust was built following the approach proposed in \cite{Malik:2023mnx}: the BPS EoS was chosen for the outer crust, and a polytropic EoS was used for the inner crust. To minimise the error introduced, the core EoS was matched to the inner crust at 0.04 fm$^{-3}$. It has been shown in \cite{Malik:2024nva} that this approximation introduces an error in the radius below 200 m.} In the same figure, we have also included the GW170817 data from the LIGO-Virgo collaboration \cite{LIGOScientific:2018cki}, together with the 2D posterior distribution in the mass-radius domain for the millisecond pulsars PSR J0030+0451 \cite{Riley:2019yda, Miller:2019cac} and PSR J0740+6620 \cite{Riley:2021pdl, Miller:2021qha} from the NICER X-ray data. 
We also illustrate the most recent mass-radius measurements from NICER for PSR J0437-4715 \cite{Choudhury:2024xbk}. For reference, the 90\% CI region obtained in a previous study using RMF and nonlinear mesonic interactions in \cite{Malik:2023mnx} is marked with a black dashed line. The very recent data predicted using the BSK24 NS EoS \cite{Goriely2013} from the GW230529 event detected by the LIGO-Virgo-KAGRA (LVK) collaboration was also included. It is  identified as the merger of two compact objects with masses of $2.5-4.5$ M$_\odot$ and $1.2-2.0$ M$_\odot$ \cite{LIGO-P2300352-v9}. This event has been interpreted by the LVK collaboration as most likely a NS-black hole merger.

{All sets except z and y seem to agree with the observations of NICER and GW170817. Set z predicts the largest radii, followed by set y. The two sets are in tension with the results derived from GW230529 using the BSK24 EoS, and set z also with GW170817. On the other hand, the three sets x, xz and xyz are consistent with the latter data, in particular the M-R 1$\sigma$ distribution obtained for GW230529 is completely within the distributions of these sets. As expected, sets xz and xyz span a larger region of the M-R diagram compared to set x, predicting slightly larger radii for the low-mass stars and the opposite for the massive stars. The high-mass behavior is dictated by the presence of the term $\omega^4$, i.e. z, which  {softens} the EOS at large densities.} 

{Compared to other approaches, as in \cite{Malik:2023mnx}, the three sets x, xz and xyz are compatible with RMF-NL, although they cover a narrower range. Note that the M-R range associated with these sets also slightly match the { 68\% CI} of PSR J0437 - 4715. In addition, we performed an inference analysis for the xz set, incorporating all the constraints defined above along with the recent data obtained by NICER for  PSR J0437 - 4715 \cite{Choudhury:2024xbk}, which showed a slight effect on the minimum of the M-R posterior, resulting in an average decrease of the radius upper limit about 0.1 km for masses ranging below $\sim$ 1.6 $M_\odot$, and the radius lower limit above $\ \sim$1.6$M_\odot$, as shown by the thick black solid line.   We have also compared the evidence of the xyz model without and with the J0437 constraint and obtained $\ln K_{\rm{xyz, xyz} J0437}=1.97$, which seems to indicate that  {there may be some tension between the new NICER data and the older observational data, and/or  that the CMF model with the nonlinear terms is more favorable for describing the old data than the old data together with the new NICER data}, {meaning that a new interaction term  or new degrees of freedom might need to be added to the model.}}

In the right panel we have plotted the mass in terms of tidal deformability for all sets, including the existing constraints from GW170817 and those derived using the BSK24 EoS from the recent GW230529 event, and for the latter constraint both the 1$\sigma$ and the 2$\sigma$ CI are shown. {Sets z and y are only marginally compatible with GW170817, but compatible with GW230529. The other three sets describe both events well. Note that the recent PSR J0437 - 4715 does not have a strong effect on the radius, but it is noticeable the effect on the lower bound of the tidal deformability distribution. }

\begin{figure}
    \centering
    \includegraphics[width=1\linewidth]{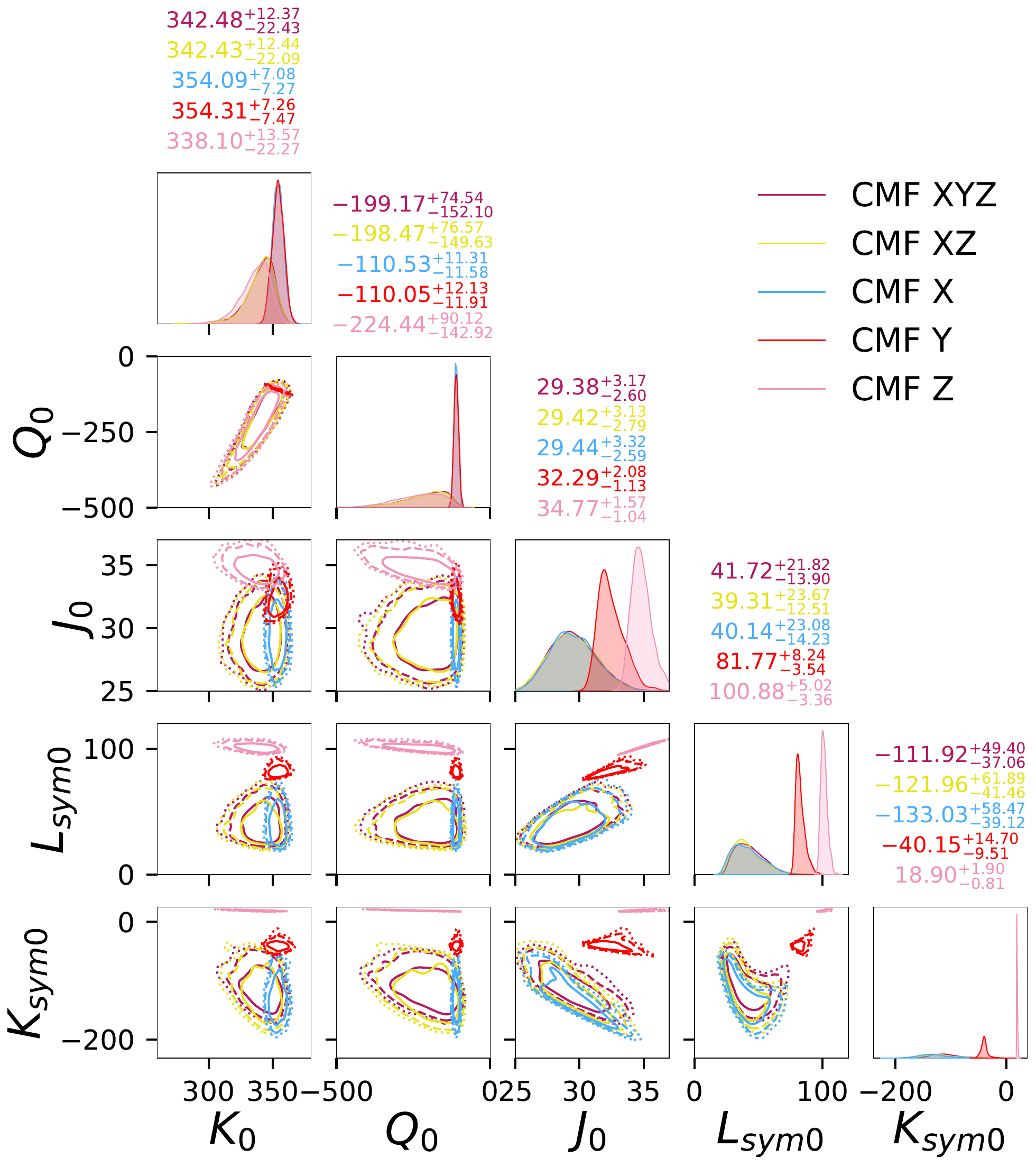}
    \caption{Corner plots for all sets of CMF posteriors using the Bayesian approach. The marginalized posterior distributions of the nuclear-matter parameters and specific NS characteristics are displayed on the diagonal. Confidence ellipses for the two-dimensional posterior distributions are depicted with intervals of $1\sigma$, $2\sigma$, and $3\sigma$ on the off-diagonal.On the diagonals,  the uncertainties are reported at a confidence level of 90$\%$.}
    \label{fig:nmp}
\end{figure}

Next, we analyze the nuclear matter properties associated with the {five} sets. In Fig. \ref{fig:symeng}, the symmetry energy is plotted as a function of baryonic density, together with the constraints from isospin analog states (IAS) determined in \cite{Danielewicz:2013upa}. All sets satisfy the IAS constraints, but set z clearly predicts a symmetry energy {that increases with density much faster than  than the other sets.}  {The symmetry energy obtained with set y is not as hard, but to get a real softening it is necessary to include the contribution of the x term ($\rho^2\omega^2$), with set x giving exactly the softest behavior}. This is confirmed by looking at the corner plot shown in Fig. \ref{fig:nmp}. {While for the sets x, xz and xyz the symmetry energy and its slope at saturation are very similar, the sets z and y have a larger symmetry energy and corresponding slope and curvature at saturation, $L_{\rm sym0}$ and $K_{\rm sym0}$, the extreme case being z for which $K_{\rm sym0}$ is always positive.}  It is also interesting to analyze the symmetric nuclear matter parameters $K_0$ and $Q_0$. {All sets have values above 300 MeV, a typical behavior of the CMF framework. $K_0$ and $Q_0$ clearly show a linear correlation. Sets x, xz and xyz have the largest  $K_0$ values necessary to allow the description of massive stars, compensating for the softer symmetry energy. Note also that the sets x and y have a very localized distribution for these two quantities. These two sets are controlled at high densities, when matter is more symmetric, by the behavior of the $\rho$ meson, which is proportional to the neutron-proton asymmetry and is weakened by the nonlinear terms. }

Recently, it has been frequently discussed in the literature how a phase transition to quark matter could be identified \cite{Annala2019,Annala:2021gom,Altiparmak:2022bke,Somasundaram:2022ztm,Fujimoto:2022ohj,Raithel:2022efm,Essick:2023fso}. These studies use agnostic descriptions of the EoS, both parametric and non-parametric, and discuss quantities that could reflect a first order phase transition, such as, the speed of sound $c_s$, the polytropic index $\gamma=\frac{d\ln p}{d\ln \epsilon}$ \cite{Annala2019}, where $p$ and $\epsilon$ are respectively the pressure and the energy density,  the trace anomaly $\Delta=1/3-p/\epsilon$,  which approaches zero in the conformal limit \cite{Fujimoto:2022ohj}, or other quantities derived from these ones. An example of the latter is the quantity $d_c=\sqrt{\Delta^2+\Delta^{'2}}$, where  $\Delta'= c_s^2 \, \left(1/\gamma-1\right)$ is the logarithmic derivative of $\Delta$ \cite{Annala:2023cwx}. In the conformal limit, $c_s^2$ and $\gamma$ approach respectively 1/3 and 1. It was   was proposed that $d_c\lesssim 0.2$ would identify the proximity of the conformal limit since for this to happen both $\Delta$ and its derivative should be small \cite{Annala:2023cwx}. Since  it is expected that quark matter shows an approximate conformal symmetry, a small value of $d_c$ could {uniquely} identify the presence of quark matter.

\begin{figure}
    \centering
    \includegraphics[width=0.99\linewidth]{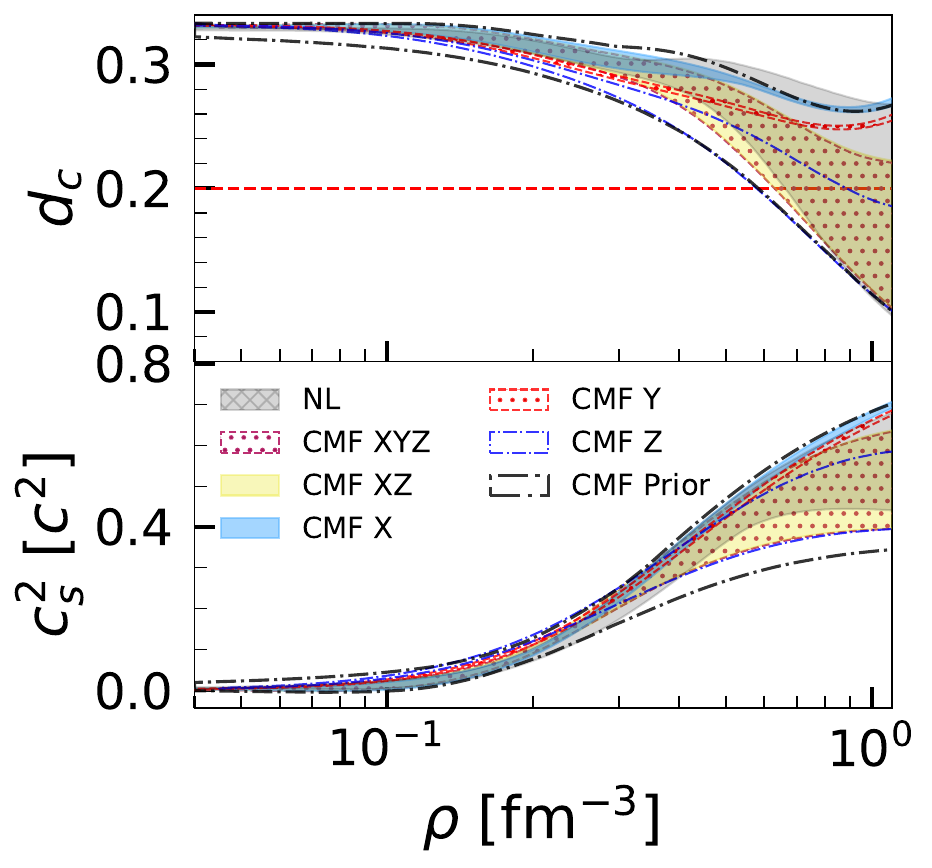}
    \caption{90\% confidence intervals are shown for the square of the speed of sound $c_s^2$ (bottom) and for $d_c$ (top)  as a function of  $\rho$ across all  CMF sets. $d_c=\sqrt{\Delta^2+ {\Delta}^{'2}}$, where  $\Delta'= c_s^2 \, \left(1/\gamma-1\right)$, is the logarithmic derivative of $\Delta=1/3-P/\epsilon$ with respect to the energy density, which approaches zero in the conformal limit \cite{Annala:2023cwx}. RMF results from \cite{Malik:2023mnx} are included as a gray hatched band. The Thick dash-dotted line delineates the xyz prior.}
    \label{fig:dc_cs2}
\end{figure}

In Fig. \ref{fig:dc_cs2} the speed of sound squared and the quantity $d_c$ are plotted for all sets. {Sets x and y have a rather narrow distribution for both quantities: at high densities both show the highest values of the speed of sound squared, approaching values $\sim 0.7$ at densities of the order of seven times  {$\rho_0$}, and values of $d_c$ well above 0.2. This behavior of both sets could be expected considering that very restricted values of $K_0$ and $Q_0$ both take, as discussed above. On the other hand, set z shows the smallest values of the speed of sound squared, approaching values $\sim 0.4-0.55$ at densities of the order of seven times  {$\rho_0$}, but also the largest values in the range of densities $\rho_0-2\rho_0$. It also predicts the smallest $d_c$ values, which fall below 0.2 at $\sim 4\rho_0$. Sets xz and xyz mix the behavior of x and z and, therefore show an intermediate behavior, being, however, completely  controlled by the z behavior at high densities. { These results suggest that the properties proposed in \cite{Annala:2023cwx} as identifying deconfined matter are not absolutely unique, because there are models of nuclear matter that do not include deconfinement but may have similar properties, and the CMF model is an example. {The term $\omega^4$ dictates this behavior.}}}

{For reference, we also include RMF results for both quantities \cite{Malik:2023mnx}. The main difference are the possible smaller values of the speed of sound at high densities  and smaller $d_c$ at intermediate densities for sets including z, reflecting the chiral symmetry of the model.}

\textbf{\textit{Conclusions-}}
 We have  used a Bayesian inference approach to study  how the    different chiral invariant vector self-interacting terms could affect the description of  {dense} nuclear matter in the framework of the nucleonic version of the Chiral Mean Field (CMF) model \cite{Dexheimer:2008ax}. Different combinations of the three  main vector self-interacting terms were considered  {and a combination of standard and new nuclear physics and astrophysics data was implemented.}

{It was shown that it is important to consider the {vector self-interacting terms  with the most impact on the equation of state to successfully satisfy observations and nuclear matter constraints from $\chi$EFT calculations, in particular,  it is necessary to consider the  $\omega\rho$ mixing term {(named x)}  together with the quartic $\omega$ term {(named z)}.  An independent contribution of the terms x and z, or the terms x, y and z allows a satisfactory description of all constraints, the overall effect of term y ($\rho^4$) being very small. In particular,  the combination of  x and z} terms are important to make the CMF model compatible with very recent NS observations, the GW230529 event by the LVK collaboration \cite{LIGO-P2300352-v9} and the mass-radius prediction for the pulsar PSR J0437-4715 by NICER \cite{Choudhury:2024xbk}, by shifting the mass-radius distributions to lower radii and the mass-tidal deformability distributions to lower tidal deformabilities.  These two terms have often been used in a relativistic mean-field description of nuclear matter to control the density dependence of  the symmetry energy and the high-density behavior of the energy density \cite{Fattoyev:2010rx}. The compatibility with the $\chi$EFT constraint implies a soft symmetry energy, with values of {its} slope and curvature at saturation of the order of 40 MeV and -110 to -130 MeV, respectively. As a consequence, the incompressibility was found to be quite high, $\sim 300-350$ MeV (but still within the experimental range - see review \cite{Stone:2014wza}). {Calculating the Bayes factor it was shown that the present model does not favor the inclusion in the inference process of the PSR J0437-4715 data, possibly indicating the model is incomplete, or that there is some tension between the new and the older data.}

{We have  {also} analyzed the behavior of the constrained model with respect to the conformal limit, following the study developed in \cite{Annala:2023cwx}. It was shown that the vector self-interacting terms have a noticeable effect both on the speed of sound, which increases monotonically with the baryon density for all combinations, and on the trace anomaly related quantity $d_c$ defined in \cite{Annala:2023cwx}, which may indicate an approach to conformality when taking values smaller than 0.2. This limit was not reached when the terms x and y ($\rho^2\omega^2$ and $\rho^4$) were considered alone, but with the other vector interaction combinations, including the term  z ($\omega^4$), it is possible to predict a value of $d_c<0.2$ above $\sim 4\rho_0$, although the present CMF model is a hadronic model that does not include a phase transition to {matter with hyperons or deconfined quark matter}, only chiral symmetry restoration. We conclude that $d_c<0.2$ is not an unambiguous indication of a deconfinement phase transition.} {Note, however, that we constrain our model imposing  nuclear matter properties at saturation, besides the  $\chi$EFT, pQCD and astrophysical constraints imposed in \cite{Annala:2023cwx}.  As shown in Fig. \ref{fig:dc_cs2},  assuming reasonable values for the symmetric nuclear matter properties at saturation,  the model does not have much freedom left to allow the EOS, and in particular $d_c$, to exhibit a behavior similar to that found in \cite{Annala:2023cwx}.}

\textbf{\textit{Acknowledgments-}}
We thank Rajesh Kumar and Mateus Pelicar for interesting discussions. This work was partially supported by national funds from FCT (Fundação para a Ciência e a Tecnologia, I.P, Portugal) under projects 
UIDB/04564/2020 and UIDP/04564/2020, with DOI identifiers 10.54499/UIDB/04564/2020 and 10.54499/UIDP/04564/2020, respectively, and the project 2022.06460.PTDC with the associated DOI identifier 10.54499/2022.06460.PTDC. The authors acknowledge the Laboratory for Advanced Computing at the University of Coimbra for providing {HPC} resources that have contributed to the research results reported within this paper, URL: \hyperlink{https://www.uc.pt/lca}{https://www.uc.pt/lca}. V. D. acknowledges support from the Fulbright U.S. Scholar Program and the National Science Foundation under grants PHY1748621, MUSES OAC-2103680, and NP3M PHY-2116686. 

%
\end{document}